\documentclass[trackchanges,twocolumn,twocolappendix]{aastex7}
\usepackage{url}
\usepackage{multirow}

\begin{document}
\title{XRISM Observations of Accretion Disk Corona in Cyg X-2}

\author[orcid=0000-0003-2161-0361]{Misaki Mizumoto}
\email[show]{mizumoto-m@fukuoka-edu.ac.jp}  
\affiliation{Science Education Research Unit, University of Teacher Education Fukuoka, Munakata, Fukuoka 811-4192, Japan}

\author[orcid=0000-0001-6314-5897]{Hiromitsu Takahashi}
\email{hirotaka@astro.hiroshima-u.ac.jp}  
\affiliation{Department of Physics, Hiroshima University, Higashi-Hiroshima, Hiroshima 739-8526, Japan}

\author[orcid=0000-0001-9735-4873]{Ehud Behar}
\email{behar@physics.technion.ac.il}  
\affiliation{Department of Physics, Technion, Technion City, Haifa 3200003, Israel}

\author[orcid=0000-0003-2704-599X]{Rozenn Boissay-Malaquin}
\email{rozennbm@umbc.edu}  
\affiliation{Center for Space Science and Technology, University of Maryland, Baltimore County (UMBC), Baltimore, MD 21250, USA}
\affiliation{NASA / Goddard Space Flight Center, Greenbelt, MD 20771, USA}

\author[orcid=0000-0002-5466-3817]{Lia Corrales}
\email{hliac@umich.edu}  
\affiliation{Department of Astronomy, University of Michigan, MI 48109, USA}

\author[orcid=0000-0001-8470-749X]{Elisa Costantini}
\email{e.costantini@sron.nl}  
\affiliation{SRON Netherlands Institute for Space Research, Niels Bohrweg 4, 2333 CA Leiden, The Netherlands}

\author[orcid=0000-0001-7796-4279]{Maria Diaz-Trigo}
\email{mdiaztri@eso.org}  
\affiliation{ESO, Karl-Schwarzschild-Strasse 2, 85748, Garching bei M\"{u}nchen, Germany}

\author[orcid=0000-0002-3031-2326]{Eric D.~Miller}
\email{milleric@mit.edu}  
\affiliation{Kavli Institute for Astrophysics and Space Research, Massachusetts Institute of Technology, Cambridge, MA 02139, USA}

\author[orcid=0000-0003-2869-7682]{Jon M.~Miller}
\email{jonmm@umich.edu}  
\affiliation{Department of Astronomy, University of Michigan, MI 48109, USA}

\correspondingauthor{Misaki Mizumoto}

\begin{abstract}
We present the high-resolution X-ray spectrum of the Z-source Cygnus X-2, obtained with X-Ray Imaging and Spectroscopy Mission (XRISM). The observations have enabled a precise characterization of the Fe-K emission lines from the accretion disk corona (ADC) and a possible detection of an ultra-fast outflow (UFO). The ADC component has at least two distinct regions. The Fe-K emission lines are remarkably broad, with a velocity dispersion of $\sigma_v=2300^{+900}_{-700}$~km~s$^{-1}$. This can be larger than what was observed in previous Chandra observations ($1120\pm870$~km~s$^{-1}$) and recent XRISM observations of the similar Z-source GX 340+0 (360--800~km~s$^{-1}$). Furthermore, we marginally detect a blueshifted absorption feature, which we identify as either \ion{Fe}{26} or \ion{Fe}{25}, with outflow velocities of $-0.29c$ or $-0.26c$, respectively. The mass loss rate is $\dot{M}\sim10^{-8}M_\odot\,\mathrm{yr}^{-1}$. Our findings suggest that the ADC in Cyg X-2 has a kinematically active environment with a high degree of turbulence and kinetic energy. 

\end{abstract}


\section{introduction}

Cygnus X-2 is one of the brightest known low-mass X-ray binaries (LMXBs) and a rare case among persistent LMXBs in which the secondary star is easily observed.
Cygnus X-2 is known to contain a neutron star because type I X-ray
bursts have been observed \citep{kahn84,kuulkers95,wijnands96,smale98}. The neutron star is believed to be accreting mass from
its companion at a near Eddington rate \citep{smale98}. 
Its relatively bright optical counterpart, V1341 Cyg \citep{giacconi67}, allows for precise spectroscopic and photometric observations.
\citet{orosz99} found an inclination of $62.5\pm4$~deg, and NS and companion masses of $1.78\pm0.23\,M_\odot$ and $0.60\pm0.13\,M_\odot$, respectively, with a distance of $7.2\pm1.1$~kpc.
More recently, \citet{ding21} used Gaia EDR3 data
to calculate a distance of $11.3^{+0.9}_{-0.8}$~kpc. We use this distance in this paper.

As one of the brightest and most persistent NS-LMXBs, Cygnus X-2 is classified as a ``Z-source''. This classification arises from the characteristic ``Z''-shaped pattern that such sources trace over hours to days in X-ray color-color and hardness-intensity diagrams \citep{hasinger1989}. The Z-sources represent the most luminous class of NS-LMXBs, accreting at rates near the Eddington limit. The Z-track itself is composed of three distinct segments: the Horizontal Branch (HB), the Normal Branch (NB), and the Flaring Branch (FB). The source's position on this track is a direct tracer of the physical state of the inner accretion flow. While motion along the track is broadly driven by the mass accretion rate, the relationship is complex. Current physical models suggest that the NB corresponds to a phase of increasing mass accretion rate, whereas the HB is a state dominated by intense radiation pressure from the neutron star that may disrupt the inner disk and launch jets, and the FB is characterized by unstable thermonuclear burning on the stellar surface (e.g., \citealt{church04, BC2010}). 

At high accretion rates, the environment immediately surrounding the neutron star is expected to be complex, consisting of a hot, X-ray-illuminated Accretion Disk Corona (ADC; e.g., \citealt{church04}) and  disk winds (e.g., \citealt{XRISM_GX13}). The Fe-K band (6--7~keV) offers one of the most powerful diagnostics for studying this plasma. The emission lines from highly ionized iron are sensitive probes of the physical conditions of the emitting gas.
Furthermore, the profiles of these lines, accessible only through high-resolution spectroscopy, encode the kinematics and geometry of the inner accretion flow. Line broadening can reveal turbulence and rotational velocity, while energy shifts can trace inflows, outflows (winds), and gravitational redshifts.

Previous observations of Cyg X-2 with Chandra/HETG have revealed broadened Fe-K lines from the ADC \citep{schulz09}, but the limited signal-to-noise ratio prevented a detailed characterization of the line kinematics and the search for more subtle features. The unprecedented spectral resolution and sensitivity of the Resolve instrument \citep{kelley2025,resolve} aboard the X-Ray Imaging and Spectroscopy Mission (XRISM) satellite \citep{XRISM} now allow us to dissect these line profiles in detail, providing a new window into the dynamics of the ADC and the launching of winds from near-Eddington accretion flows.

Here, we present the results of the high-resolution X-ray spectroscopic observations of Cyg X-2 by XRISM. Using the high-resolution Resolve spectrometer, we report the iron K emission lines originating from ADC and the possible ultra-fast outflow (UFO) absorption lines. This paper is organized as follows. Section 2 describes the XRISM observation and data reduction process. In Section 3, we present the spectral analysis and model fitting of the XRISM data. Section 4 is dedicated to a discussion of our findings. Finally, in Section 5, we summarize our conclusions.
The errors correspond to 1$\sigma$ statistical uncertainty throughout this paper.


\section{Observation and data reduction}
\begin{figure}[h]
   \centering
    \includegraphics[width=0.95\columnwidth]{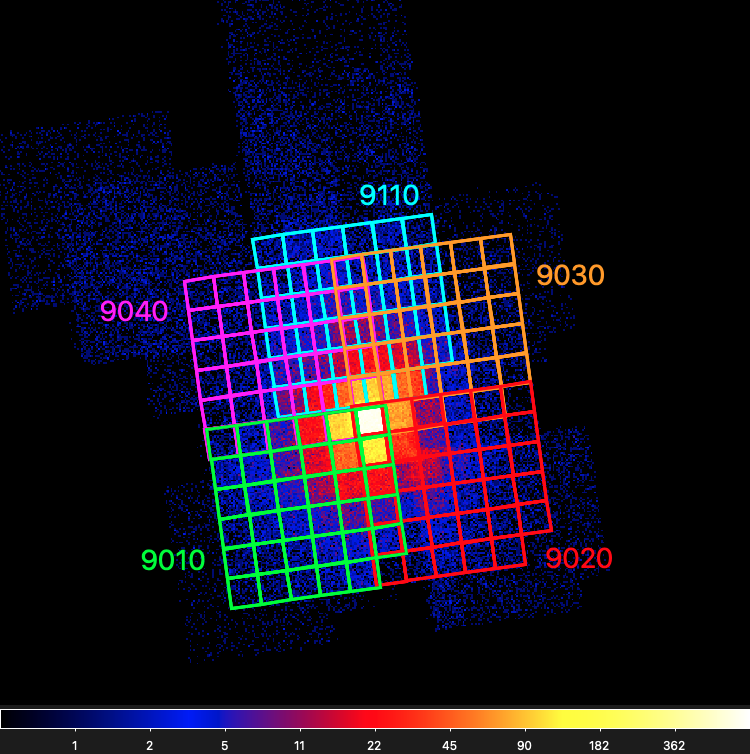}
            \caption{The mosaic image in Resolve. The five fields of view used in the spectral analysis are shown by color.}
         \label{fig:ds9}
   \end{figure}

\begin{figure}[ht]
   \centering
    \includegraphics[width=0.9\columnwidth]{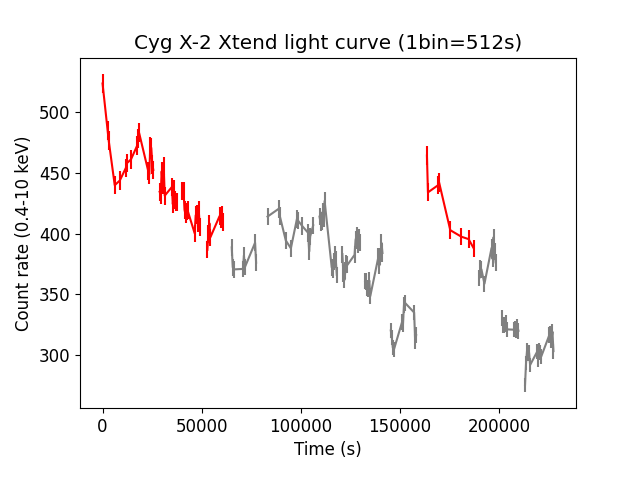}
    \includegraphics[width=0.9\columnwidth]{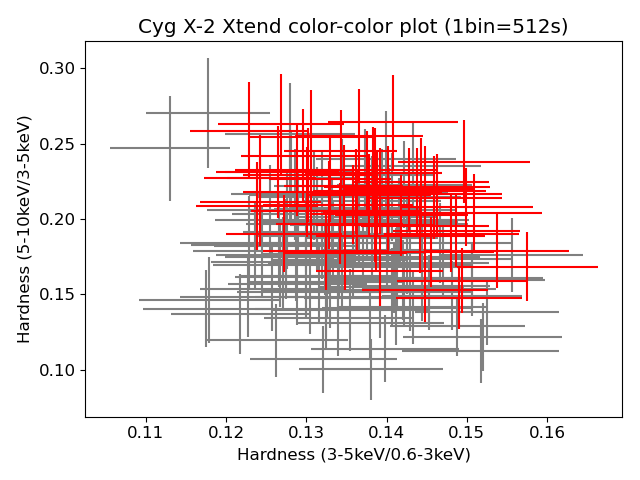}

            \caption{(Upper) Xtend light curve with a time bin of 512~s. The data we used for the spectral analysis is shown in red. (Lower) Color-color plot based on the Xtend light curve.}
         \label{fig:ltcrv}
   \end{figure}

XRISM observed Cyg X-2 as part of the in-flight calibration program, with the main aim of studying the off-axis point spread function \citep{miller2021}. 
The observation was performed from 2024-04-20 to 2024-04-23 (UTC), with different offset positions. Each exposure time is relatively short.
The observation log is listed in Table \ref{tab:obs}, and the mosaic image in Resolve is shown in Figure \ref{fig:ds9}.
We used five of the fourteen total observations, selecting those with small pointing offsets.
Figure \ref{fig:ltcrv} shows the Xtend light curve and a color–color plot. The count rate varies by less than a factor of two. For the data used in the spectral analysis, the variability is at most about 20\%. The color-color plot indicates that the source was observed in the upper-right region of its ``Z''-track, corresponding to the period when the HB and NB overlap. We note that it is not on the Flaring Branch FB since no short-term variability is seen, {and that
the derived parameters of the blackbody emission (see \S3.1 for the temperature and \S4.1 for the radius) are consistent with the ones at the transition of HB and NB (e.g., \citealt{BC2010}).}

\begin{deluxetable*}{cccccc}
\tablewidth{0pt}
\tablecaption{Observation Data \label{tab:obs}}
\tablehead{
\colhead{OBSID} & \colhead{Name\tablenotemark{*}} & \colhead{RA} & \colhead{Dec} & \colhead{Time} & \colhead{Resolve Exposure (s)}
}
\startdata
100009010 & Cyg\_X-2\_1p8\_45   & 21 44 46.73 & +38 17 50.8 & 2024-04-20 09:04:04.0 & 16524 \\
100009020 & Cyg\_X-2\_1p8\_135  & 21 44 34.14 & +38 18 14.4 & 2024-04-20 17:29:04.0 & 8072  \\
100009030 & Cyg\_X-2\_1p8\_225  & 21 44 35.88 & +38 20 45.6 & 2024-04-20 20:42:04.0 & 8038  \\
100009040 & Cyg\_X-2\_1p8\_315  & 21 44 48.70 & +38 20 23.0 & 2024-04-20 23:54:04.0 & 7505  \\
100009110 & Cyg\_X-2\_1p8\_270  & 21 44 42.69 & +38 21 06.0 & 2024-04-22 06:27:04.0 & 13598 \\
\enddata
\tablenotetext{*}{The name means offset and offset angle. For example, ``Cyg\_X-2\_1p8\_45'' means the offset of 1.8 arcsec with an offset angle of 45 deg.}
\end{deluxetable*}

XRISM has two instruments: Resolve and Xtend \citep{noda2025,uchida2025}.
The observations were performed with a Neutral Density (ND) filter, whose transmittance is 25\%, for Resolve (with the gate valve closed) and {using the full-window + burst mode for Xtend}. 
HEASOFT v6.35.1 was used for the data reduction. 
Data were processed with the pre-pipeline version ``005\_001.20Jun2024\_Build8.012'', the pipeline script ``03.00.013.009'', and the CALDB version 20240815.

\begin{figure}[ht]
   \centering
    \includegraphics[width=0.95\columnwidth]{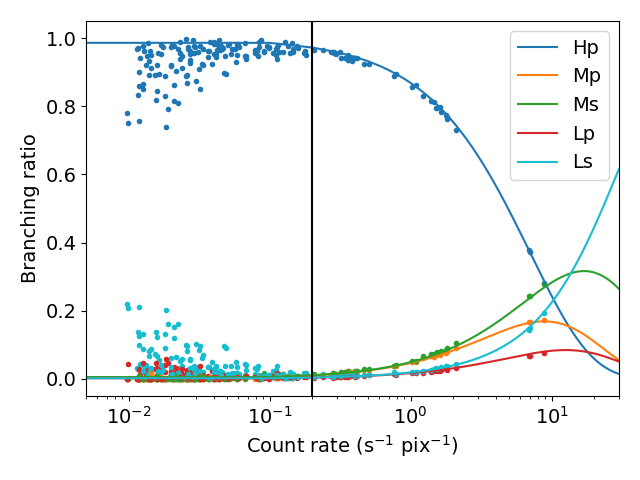}
            \caption{The event branching ratio against the count rate in each pixel. The model line is overplotted, calculated under the assumption that the events follow the Poisson distribution. We use the pixels with count rates larger than 0.2 cts~s$^{-1}$~pix$^{-1}$ (vertical line). Hp: High-resolution primary, Mp: Mid-resolution primary, Ms: Mid-resolution secondary, Lp: Low-resolution Primary, and Ls: Low-resolution secondary.}
         \label{fig:branch_all}
   \end{figure}

The routines used to perform the data reduction and extraction of the products followed the steps presented in the ``XRISM ABC Guide''\footnote{\url{https://heasarc.gsfc.nasa.gov/docs/xrism/analysis/abc_guide/xrism_abc.pdf}}, but we have applied additional reduction.
Some of the events detected with Resolve are contaminated by false events known as anomalous Low-resolution secondary (Ls) events (see ``Things to watch out for with XRISM data processing and analysis (TTWOF)''\footnote{\url{https://heasarc.gsfc.nasa.gov/docs/xrism/analysis/ttwof/index.html}}). 
The rate of these events does not depend on the source count rate. Therefore, when the source count rate is relatively high, the fraction of false events is negligible. On the other hand, when the count rate is low, the fraction of anomalous Ls events increases, distorting the intrinsic event branching ratio. Figure \ref{fig:branch_all} shows the dependence of the event branching ratio on the pixel count rate. When the count rate is sufficiently low, the Hp fraction is expected to be nearly 100\%, but in reality, it becomes smaller due to the presence of anomalous Ls events. To avoid this effect, we do not use pixels with count rates lower than 0.2 cts~s$^{-1}$~pix$^{-1}$.
The maximum count rate per pixel in Resolve including all the event grade is 8.8~cts~s$^{-1}$~pix$^{-1}$. In this count rate, change of the spectroscopic capability due to high-count-rate effects are not expected \citep{mizumoto2025a,mizumoto2025b}.
In the Xtend data reduction, the source region was extracted using an annular region with an inner radius of 10 pixels (17.68 arcsec) and an outer radius of 50 pixels (88.4 arcsec), to avoid the pile-up effect (\citealt{yoneyama2024}; XRISM ABC Guide).
Backgrounds in both Resolve and Xtend can be ignored, since their level is much lower than the source level.
A Response Matrix File (RMF)  was created using rslmkrmf (with the ``L'' option, in which Gaussian core, exponential tail, Si K$\alpha$ instrumental line, and escape peak are taken into account) for Resolve and xtdrmf for Xtend.
Ancillary Response Files (ARF) were generated using xaarfgen, assuming a point-like source at Cyg X-2's coordinates as input.
The five observations have been combined. Specifically, the source spectra were added using \texttt{mathpha}. The RMFs and ARFs were first merged using \texttt{ftmarfrmf}, and then summed using \texttt{ftaddrmf} with weights of the exposure time.
We use only the Hp grade for Resolve. The resulting Hp count rate is 6.45~cts~s$^{-1}$.
Throughout the paper, the model fitting was performed without the binning data and with the C statistics \citep{cash1979}, while in some figures the binning of the energy for the plot has been changed for the sake of clarity.

\section{Results}
\subsection{Continuum modeling}\label{sec3.1}
\begin{figure}[h]
   \centering
    \includegraphics[width=0.95\columnwidth]{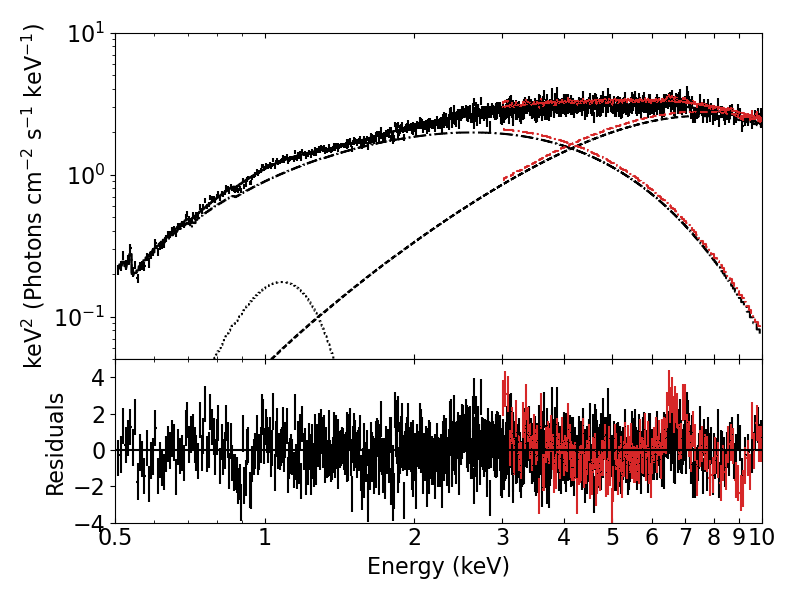}
    \includegraphics[width=0.95\columnwidth]{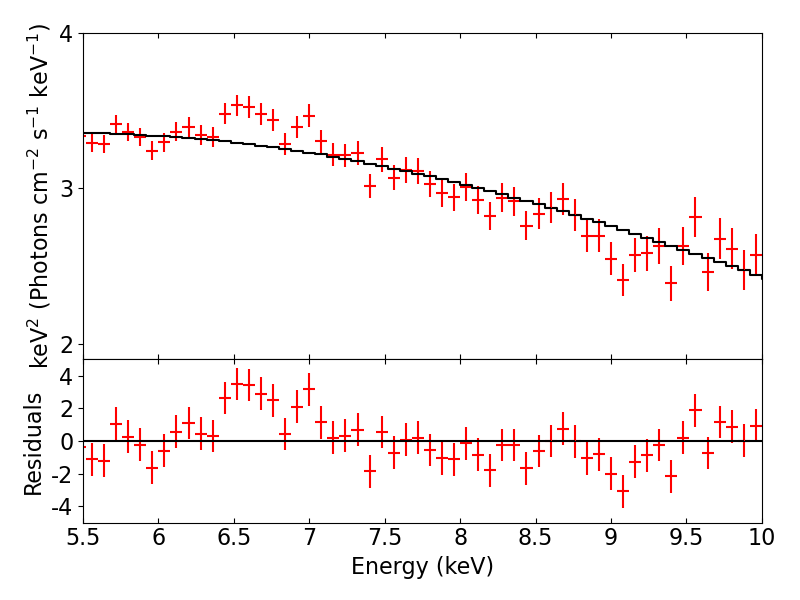}
            \caption{{Continuum fitting. (Upper) The upper panel shows the $\nu F_\nu$ plot, and the lower panel shows (data$-$model)/error. The Xtend and Resolve data are shown in black and red, respectively.
            (Lower) Same as the upper one, but only Resolve data with more sparse binning (80~eV).}}
         \label{fig:continuum}
   \end{figure}

First the continuum model was constrained with the simultaneous fitting of the 0.5--10 keV Xtend spectrum and the 3--10 keV Resolve spectrum.
We do not use the Resolve data below 3 keV since residual from the Xtend data cannot be ignorable.
We start with two thermal components ({\tt diskbb}+{\tt bbodyrad}) from the accretion disk and blackbody emission from the NS or boundary region. 
To explain the residuals around 1~keV, 
a positive Gaussian was introduced phenomenologically \citep{BC2011}.
The foreground absorption is modeled by {\tt tbabs}, with the abundance of \citet{wilm}.
The fitting results are shown in Figure \ref{fig:continuum} and Table \ref{tab:continuum}. 
The cross normalization between Resolve and Xtend, 1.074, is relatively higher than the reported one \citep{xrism_xcal}. Several factors may contribute to this discrepancy. First, both the Resolve and Xtend spectra were summed over multiple observations. Second, the Resolve data have offset pointings, capturing only one side of the PSF wing, which can introduce uncertainties in the effective area calculation. Third, variations in the pixel-by-pixel calibration may also affect the cross normalization (see RSL-3 and RSL-4 in TTWOF).
This paper does not further discuss it.
The unabsorbed flux within 1--1000 Ryd is $1.2\times10^{-8}$~erg~cm$^{-2}$~s$^{-1}$ (Xtend). With the distance of $11.3$~kpc, the luminosity is $1.9\times10^{38}$~erg~s$^{-1}$ and the Eddington ratio is 0.9.

\begin{deluxetable}{ccc}
\tablewidth{0pt}
\tablecaption{Continuum model \label{tab:continuum}}
\tablehead{
\colhead{Model} & \colhead{Parameter} & \colhead{Value} 
}
\startdata
tbabs & $N_\mathrm{H}$ (cm$^{-2}$) & $1.52\pm0.04 \times 10^{21}$\\
bbodyrad & $kT$ (keV) & $1.92\pm0.0$2\\
 & norm & $38.9\pm2.4$\\
diskbb & $kT_\mathrm{in}$\,(keV) & $1.076\pm0.018$\\
& norm & $225\pm13$\\
gaussian & $E_c$ (keV) & $0.992\pm0.019$ \\
& $\sigma$ (keV) & $0.196\pm0.016$\\
& norm & $0.101\pm0.012$ \\
Cross norm. & Xtend & 1 (fix) \\
& Resolve & $1.074\pm0.004$ \\
\enddata
\end{deluxetable}

\subsection{Phenomenological fitting}

\begin{figure}[h]
   \centering
    \includegraphics[width=0.95\columnwidth]{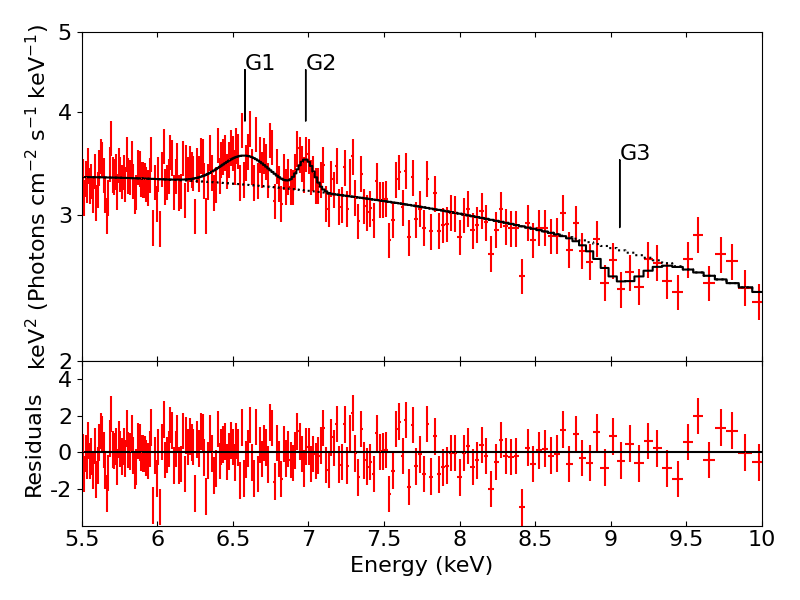}
            \caption{The Resolve model fitting with two positive Gaussian at the Fe-K band and a negative Gaussian in absorption for the UFO. The dotted line shows the continuum.}
         \label{fig:phenomenological}
   \end{figure}

As a result of the continuum modeling, some residuals are seen in the Resolve spectrum. First, the two emission line-like residuals are seen at 6.6 keV and 7.0 keV (see the lower panel of Figure \ref{fig:continuum}). Next, an absorption line is seen at 9 keV. Three Gaussian ($G_{1,2,3}$) are introduced for phenomenological fitting. The best-fit results are shown in Figure \ref{fig:phenomenological} and Table \ref{tab:gauss}. 
The equivalent widths (EW) are also listed.
The C statistic is 14216.86 with degrees of freedom (dof) of 13989.
{The $\Delta C$ values in Table \ref{tab:gauss} represent the improvement in the C-statistic between the best-fit model and the one when the specific component is removed and the remaining  parameters are re-fitted. }
The null probability of the significance of each Gaussian based on Akaike Information Criteria (AIC; \citealt{akaike74,tan12})  is also calculated, {based on the $\Delta C$ values.}

The two positive Gaussian at the Fe-K band have also been reported in the Chandra/HETGS observation in 2007 \citep{schulz09}, which can come from the ADC.
First, $G_2$ at 7.0 keV can be identified as the \ion{Fe}{26} line. Since $G_1$ has a broader line width ($\sigma_v=6800^{+1800}_{-1400}$~km~s$^{-1}$) than $G_2$ ($\sigma_v=2300^{+900}_{-700}$~km~s$^{-1}$) and its energy is lower than the \ion{Fe}{25} line at 6.63--6.70~keV, the $G_1$ can be a superposition of the \ion{Fe}{25} and lower-ionized Fe lines.
The absorption line at 9 keV ($G_3$) is possibly detected. It can be attributed to an UltraFast Outflow (UFO). The wind velocity is $v=-0.26c$ if the line is \ion{Fe}{26}, or $v=-0.29c$  if \ion{Fe}{25}.

\begin{deluxetable}{ccccccc}
\tablewidth{0pt}
\tablecaption{Gaussian modeling \label{tab:gauss}}
\tablehead{
\colhead{} &\colhead{$E_c$ (keV)} & \colhead{$\sigma$ (keV)} & \colhead{norm ($\times10^{-4}$)} & \colhead{EW (eV)}& $\Delta C/\Delta \mathrm{dof}$ & $P_\mathrm{AIC}$
}
\startdata
$G_1$ & $6.58\pm0.03$ & $0.15_{-0.03}^{+0.04}$ & $23\pm4$  & $32\pm6$ &$-50.83/3$ & $1.8\times10^{-10}$  \\
$G_2$ & $6.982_{-0.026}^{+0.015}$ & $0.053_{-0.017}^{+0.021}$ & $7\pm2$& 
$11\pm3$ &
$-17.09/3$ & 0.004 \\
$G_3$ & $9.06_{-0.05}^{+0.06}$ & $0.13_{-0.05}^{+0.06}$ & $-9\pm3$ &
$-28\pm9$ &
$-12.75/3$ & 0.034\\
\enddata
\end{deluxetable}

\subsection{ Modeling with \texttt{XSTAR}}

The ADC emission lines are modeled using the  \texttt{XSTAR} photoionization code (ver 2.59d; \citealt{xstar}). Assuming that the line labeled $G_2$ corresponds to the \ion{Fe}{26} transition, we adopted a velocity broadening of $\sigma_v = 2300$~km~s$^{-1}$ 
(\texttt{vturbi} in xstar is $\sqrt{2}\,\sigma_v = 3300$ km~s$^{-1}$). A table model  for a photoionization equilibrium (PIE) plasma was calculated. The spectral energy distribution (SED) determined from the continuum modeling (\S\ref{sec3.1}) was used as input.

\begin{figure}[h]
   \centering
    \includegraphics[width=0.95\columnwidth]{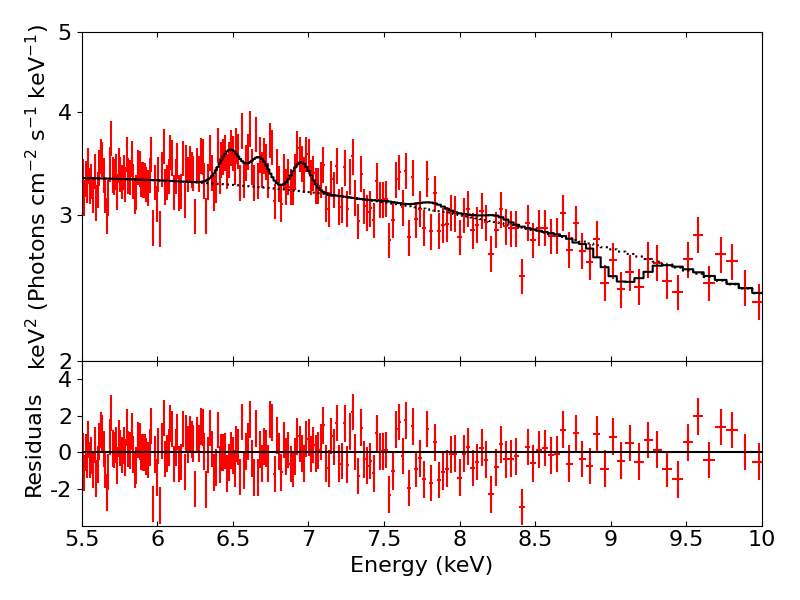}
            \caption{Same as Figure \ref{fig:phenomenological}, but fitted with two photoionized plasma components for the emission lines. The case for a covering fraction (\texttt{cfrac}) of 0.5 is shown. }
         \label{fig:cfrac50}
   \end{figure}

A single emission component could not reproduce both $G_1$ and $G_2$ simultaneously. To address this, we introduced an additional emission component with the same velocity dispersion. The best-fit model is shown in Figure~\ref{fig:cfrac50} and Table~\ref{tab:xstar}.  
The C statistic is 14204.68 with dof of 13990 (for \texttt{cfrac}=0.5, explained in the following paragraph).
First, the high-$\xi$ component produces the \ion{Fe}{26} line, which explains the $G_2$ feature. At the same time, the \ion{Fe}{25} line explains the blue half of the $G_1$ line. Next, the low-$\xi$ component creates emission lines from \ion{Fe}{18}--\ion{Fe}{20}, which explain the red side of the $G_1$ line.
In other words, the $G_2$ feature is interpreted as a blend of \ion{Fe}{25} and \ion{Fe}{18}--\ion{Fe}{20} emission lines.
The lines are too broadened to resolve the \ion{Fe}{25} triplet, unlike the XRISM observations of GX 340+0 \citep{Chakraborty2025}. The ADC components are slightly redshifted.

In \texttt{XSTAR}, the normalization of the emission component is defined such that \texttt{norm} = 1 corresponds to a source with $L = 10^{38}$~erg~s$^{-1}$ at a distance of 1~kpc. Given the luminosity of Cyg~X-2 of $1.9 \times 10^{38}$~erg~s$^{-1}$ and a distance of 11.3~kpc, the expected normalization is 0.015. We therefore fixed the normalization to this value in our modeling.

One of the parameters that govern the strength of the emission lines is the covering fraction (\texttt{cfrac}). A smaller \texttt{cfrac} allows more photons to escape, resulting in stronger line emission for a given column density ($N_{\mathrm{H}}$). Since the current data do not allow us to constrain \texttt{cfrac} independently, we performed calculations for three representative values: 0.1, 0.5, and 0.9. The ionization parameter ($\log \xi$) remained nearly constant across these models, while the required $N_{\mathrm{H}}$ increased with increasing \texttt{cfrac}, as expected. In all cases, $N_{\mathrm{H}}$ was on the order of $10^{22}$~cm$^{-2}$ or $ 10^{23}$~cm$^{-2}$.


We then proceeded to model the absorption features using the \texttt{ionabs} model \citep{ionabs}.
This model measures ion column density from absorption lines via the Voigt profile. The wind parameters are shown in Table \ref{tab:xstar2}. The absorption line can be \ion{Fe}{25} or \ion{Fe}{26}. In both cases, the line broadening is $\sigma_v\sim6800^{+3000}_{-2300}$~km~s$^{-1}$. This is larger than the thermal broadening ($
\sigma_{v,\mathrm{thermal}}=\left(2kT_\mathrm{C}/m_\mathrm{ion}\right)^{1/2}\sim80\,\mathrm{km\,s}^{-1}$, where $kT_\mathrm{C}\sim2$~keV is the Compton temperature and $m_\mathrm{ion}$ is the iron ion mass), and even larger than the  UFOs in AGN ($\sigma_v=1900$~km~s$^{-1}$ for PDS 456; \citealt{XRISM_PDS456}). 
This could be due to velocity shear within the outflow (see e.g., \citealt{mizumoto2021}) or velocity changes during the observations (see e.g., \citealt{Gu2025}).

\begin{deluxetable}{ccccc}
\tablewidth{0pt}
\tablecaption{Emission line modeling \label{tab:xstar}}
\tablehead{
\colhead{cfrac} &\colhead{$N_\mathrm{H}$ ($10^{22}$~cm$^{-2}$)} & \colhead{$\log\xi$ (cgs)} & \colhead{$z$} & \colhead{$\sigma_v$ (km s$^{-1}$)} 
}
\startdata
0.1 & $3.4_{-1.0}^{+1.3}$ & $2.04^{+0.08}_{-0.04}$ & \multirow{2}{*}{$(1.9\pm0.9)\times10^{-3}$} &  \multirow{2}{*}{2300 (fix)} \\
& $4.3\pm1.2$ & $3.74_{-0.17}^{+0.15}$ & & \\
\hline
0.5 & $3.6_{-1.0}^{+1.5}$ & $2.05^{+0.08}_{-0.05}$ & \multirow{2}{*}{$1.9_{-0.9}^{+1.0}\times10^{-3}$} &  \multirow{2}{*}{2300 (fix)}\\
& $5.0_{-1.0}^{+1.4}$ & $3.74_{-0.16}^{+0.15}$ & & \\
\hline
0.9 & $5.5_{-1.4}^{+1.5}$ & $2.08^{+0.08}_{-0.05}$ & \multirow{2}{*}{$2.4_{-0.9}^{+0.7}\times10^{-3}$} & \multirow{2}{*}{2300 (fix)}  \\
& $10_{-2}$ & $3.71\pm0.10$ & & \\
\enddata
\end{deluxetable}

\begin{deluxetable}{cccc}
\tablewidth{0pt}
\tablecaption{Absorption line modeling \label{tab:xstar2}}
\tablehead{
\colhead{} &\colhead{$N_\mathrm{ion}$ ($10^{17}$~cm$^{-2}$)} & \colhead{$\sigma_v$ (km s$^{-1}$)} & \colhead{$v/c$}  
}
\startdata
H-like Fe & $5.2_{-1.5}^{+1.3}$ & $6800_{-2300}^{+3000}$& $-0.257\pm0.004$\\
He-like Fe & $2.8_{-0.9}^{+1.0}$ & $6800_{-2300}^{+3000}$& $-0.294\pm0.005$\\
\enddata
\end{deluxetable}

\section{Discussion}

\subsection{Radii of the blackbody radiation and the disk}
The normalization of the blackbody radiation means 
$R_\mathrm{km}^2/D_{10}^2=38.9\pm2.4$, where $R_\mathrm{km}$ is the source radius in km and $D_{10}$ is the distance to the source in units of 10~kpc. With $D_{10}=1.13$, $R_\mathrm{km}=7.0 \pm 0.2$~(km).
The normalization of the disk blackbody implies $(R_\mathrm{in}/D_{10})^2\cos\theta=225\pm13$, where $R_\mathrm{in}$ is an inner disk radius in km and $\theta$ is the angle of the disk. With $D_{10}=1.13$ and $\theta=62.5$~deg \citep{orosz99}, $R_\mathrm{in}=25.0\pm0.8$~(km), or $9.7 \pm0.3\,R_g$ with the NS mass of $1.78\,M_\odot$.

\subsection{Comparison with the Chandra 2007 observations}
The two Fe-K emission lines were also reported in the Chandra/HETG observations in 2007 \citep{schulz09}.
During the observation, Cyg X-2 experienced HB, NB, and FB phases, and a model fitting was performed on the combined spectrum.
The lower energy component ($G_1$) has $E_\mathrm{c}=6.662\pm0.011$~keV with the broadening of $\sigma=0.077\pm0.016$~keV. 
The higher energy component ($G_2$)  has $E_\mathrm{c}=6.92\pm0.02$~keV with the broadening of $\sigma=0.026\pm0.012$~keV. 

First we focus on the higher energy component ($G_2$), since it may be pure \ion{Fe}{26} emission. In this component, the line shift from 2007 to 2024 is $+60\pm30$ eV (from $6.92\pm0.02$~keV to $6.982^{+0.015}_{-0.026}$~keV), and the line width in 2024 ($\sigma=0.053^{+0.021}_{-0.017}$) is twice as large as in the 2007 data ($\sigma=0.026\pm0.012$~keV).
Next, in the lower energy component ($G_1$), the line shift is $-80\pm30$ eV (from $6.662\pm0.011$~keV to $6.58\pm0.03$~keV), which is in the opposite direction to that of G2, and the line broadening is more than twice as large (from $\sigma=0.077\pm0.016$ to $\sigma=0.15^{+0.04}_{-0.03}$). 
This can be interpreted as follows: overall, the line is slightly shifted toward the positive side, as in G2, but the increased contribution from emission lines originating from lower-ionization Fe makes the line appear broader and shifted to lower energies. In fact, Figure 2 in \citet{schulz09} shows a positive residual on the low-energy side of \ion{Fe}{25}, which could be explained by the effect of low-ionization iron ions.

\subsection{Comparison with the other Z-like source}

The Fe-K emission lines are compared with the other super-Eddington NS binaries (i.e., the Z-sources), GX~340+0 and GX~5--1.
In GX~340+0, three ionization zones are required, with parameters of $\log\xi=1.7-3.7$ and $N_{\rm H}=(3-10)\times10^{22}$~cm$^{-2}$ \citep{Chakraborty2025}. They are very similar to those of Cyg X-2. The main difference lies in the line width: Cyg X-2 shows a velocity dispersion of $2300^{+900}_{-700}$~km~s$^{-1}$, significantly larger than those of GX 340+0, 360--800~km~s$^{-1}$. 
This difference may indicate that the ADC of Cyg X-2 is a more active kinematic environment with higher turbulence and kinetic energy than that of GX 340+0. This enhanced turbulence could be caused by several factors. For instance, Cyg X-2's near-Eddington accretion rate could lead to a geometrically thicker, radiation-pressure-dominated inner disk, which in turn could sustain a more turbulent and extended corona. Alternatively, the specific inclination angle of Cyg X-2 ($62.5\pm4$~deg) might provide a clearer line of sight into the most turbulent regions of the corona, which could be obscured in other sources viewed at different angles. Further theoretical modeling is needed to disentangle these effects and fully understand the origin of the observed kinematic differences among the Z-sources.

GX~5--1 has exhibited no significant Fe-K emission features (e.g., \citealt{Homan_2018}). 
Its Eddington ratio is highest ($\sim2$ times the Eddington luminosity) so that even ADC can be fully ionized and thus cannot produce any emission lines. 
Its Resolve spectrum will be shown in Hayashi et al.\ in prep.

\subsection{UFO feature}
The absorption line at 9.06~keV may originate from the UFO.
The escape radius is $2\,(v/c)^{-2}\,R_\mathrm{g}
=30\,R_\mathrm{g}$ (\ion{Fe}{26}) and $23\,R_\mathrm{g}$ (\ion{Fe}{25}), which is larger than the estimated disk inner radius, $9.7\pm0.3\,R_\mathrm{g}$.
Such UFOs have been seen in several X‐ray binaries (e.g., \citealt{king2012,miller2015,miller2016,balakrishnan2020}). 

The potential presence of such an outflow is supported by the archival NICER spectra of Cyg X-2, which also exhibit UFO-like absorption features (Takahashi et al.\ in prep). These features appear consistently across different flux states, with the line centroid shifting from $\sim7$ keV (no velocity shift) to $\sim8.8$ keV ($v/c = 0.25$) as the source flux increases. Although the statistical quality of any single NICER observation is lower than that of our XRISM data, the persistent presence of this feature in multiple observations strengthens our interpretation. 
Therefore, while the statistical significance of the XRISM data alone is moderate ($P_\mathrm{AIC}=0.034$), its combination with supporting evidence from archival NICER spectra provides a compelling case for the presence of a UFO in this source.

Assuming that the UFO feature is the \ion{Fe}{26} line, we calculate the mass loss rate. Using $N_\mathrm{ion}=5.2\times10^{17}$~cm$^{-2}$ and the iron abundance normalized to the hydrogen of $2.69\times10^{-5}$ \citep{wilm}, the hydrogen column density is 
\begin{equation}
    N_\mathrm{H}=3.9\times10^{22} \,\left(\frac{f_\mathrm{Fe26}}{0.5}\right)\,\mathrm{cm}^{-2},
\end{equation}
where $f_\mathrm{Fe26}$ is the fraction of iron ions present in the \ion{Fe}{26} state. 

The mass loss rate is
\begin{equation}
\dot{M} = 4\pi f_{\text{cov}} f_{\text{vol}} \mu m_p n R^2 v,
\label{eq:mdot_fundamental}
\end{equation}
where $f_{\text{cov}}$ is the covering factor, $f_{\text{vol}}$ is the volume filling factor, $\mu\sim1$ is the mean ionic mass in units of proton mass ($m_p$), $n$ is the number density, and $R$ is the radius of the absorber.
By substituting the term $nR^2$ using the definition of the ionization parameter ($\xi=L_\mathrm{ion}/nR^2$), the expression can be rewritten as
\begin{equation}
\dot{M} = 10^{-8} \left(\frac{f_{\text{cov}}}{0.1}\right) \left(\frac{f_{\text{vol}}}{0.1}\right) \left(\frac{v}{0.26c}\right) \left(\frac{L_{\text{ion}}/\xi}{10^{33}\,\text{cgs}} \right)M_{\odot}~\text{yr}^{-1}.
\label{eq:mdot_normalized}
\end{equation}
Here $\log\xi\sim5$ under the assumption that \ion{Fe}{26} is dominant in the iron-ion population. $f_{\text{cov}}$ and $f_{\text{vol}}$ are difficult to constrain with current data, and we assumed a representative value of 0.1 for both.

\section{Summary}
In this paper, we used the exceptional spectral resolution of the XRISM Resolve instrument above 6~keV to perform a comprehensive study of ADC and disk winds in the neutron star LMXB Cyg X-2. Our primary objectives were to characterize the Fe-K emission lines from the ADC and to search for any outflow signatures, thereby providing new insights into the kinematics of this prototypical Z-source.

The present analysis of the Resolve spectrum revealed two key findings. First, the Fe-K emission lines originating from the ADC exhibit an unusually large velocity broadening, with a velocity dispersion of $2300^{+900}_{-700}$~km~s$^{-1}$ . 
This result is in stark contrast to the much narrower line widths reported for the similar Z-source GX 340+0 \citep{Chakraborty2025} and is also broader than previously measured by Chandra in Cyg X-2 \citep{schulz09}. This kinematic difference suggests that the ADC of Cyg X-2 is a more turbulent and dynamically active region compared to other Z-sources. We speculate that this disparity could be linked to variations in the inclination angles, mass accretion rates, or the geometrical structure of the disk-corona system between these objects.

Second, we report the possible 
detection of an ultra-fast outflow (UFO) in Cyg X-2. A distinct blue-shifted absorption line was identified in the spectrum, consistent with highly ionized iron being driven away from the central source at the relativistic velocity. Its line width is as large as $\sigma_v=6800^{+3000}_{-2300}$~km~s$^{-1}$, so that there can be some velocity shear or velocity change during the observations. The mass loss rate in this wind component is estimated to be $\dot{M}\sim10^{-8}\,M_\odot\,\mathrm{yr}^{-1}$.

In conclusion, XRISM observations provide a novel and detailed view of the accretion-outflow dynamics in Cyg X-2. The combination of an exceptionally broad ADC and a newly discovered UFO presents a compelling case for a highly energetic and turbulent environment. 

\begin{acknowledgments}
This study is financially supported from JSPS KAKENHI Grant Number JP21K13958, Yamada Science Foundation (MM), NASA grants 80NSSC20K0733, 80NSSC24K1148, 80NSSC24K1774 (EB), 80GSFC24M0006 (RBM), 80NSSC18K0978, 80NSSC20K0883, 80NSSC25K7064 (LC), 80NSSC20K0737, 80NSSC24K0678 (EDM), and JSPS Core-to-Core Program (grant number: JPJSCCA20220002).
{This paper employs a list of Chandra datasets, obtained by the Chandra X-ray Observatory, contained in the Chandra Data Collection ~\dataset[DOI: 10.25574/cdc.506]{https://doi.org/10.25574/cdc.506}.}

\end{acknowledgments}

\begin{contribution}
MM was responsible for writing and submitting the manuscript.
HT, EB, RBM, LC, EC, MDT, EDM, and JMM contributed to the discussion and manuscript check.


\end{contribution}

\facility{XRISM \citep{XRISM}}

\software{HEASoft/FTOOLS \citep{heasoft,ftools}
          }


\appendix

\section{Alternative model for the Fe-K emission lines}

As discussed in the main text, the Fe-K band appears to have multiple emission line features. Although this was explained in the main text as a superposition of multiple ADC components, there is another possibility: A single broad emission line spanning 6–7 keV could be intercepted by an absorption line at 6.8 keV.
It should be noted that the Fe-K emission line in Cyg X-2 was  often interpreted as a relativistic reflection \citep{shaposhnikov2009, cackett2010}.

   \begin{figure}[b]
   \centering
    \includegraphics[width=0.95\columnwidth]{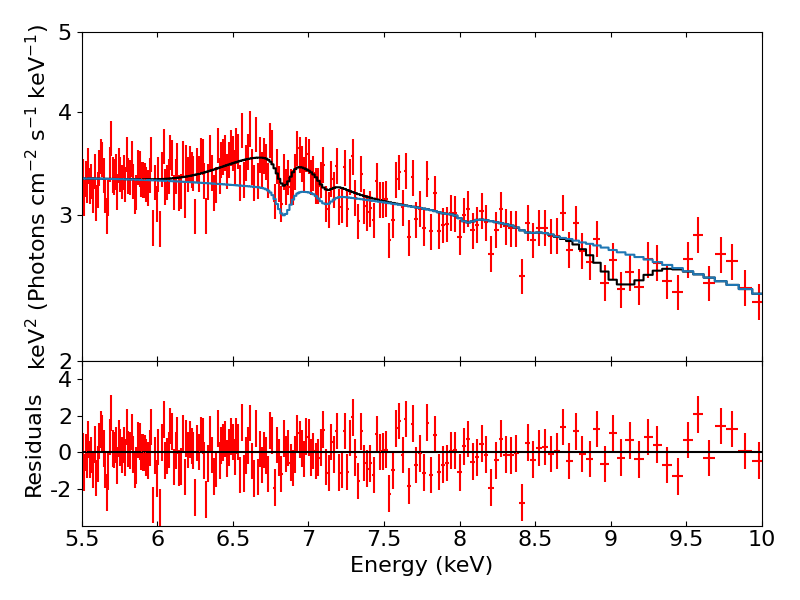}
            \caption{Same as Figure \ref{fig:phenomenological}, but one positive broad Gaussian and one blueshifted absorber are introduced to explain the 6--7~keV features. The blue line shows the continuum affected by the absorber. The 9~keV feature is explained by a negative Gaussian.}
         \label{fig:zxipcf}
   \end{figure}
Based on this scenario, we fitted the spectrum with a model composed of a continuum, a broad positive Gaussian, and a blueshifted absorber. The absorber was represented by a zxipcf model \citep{reeves2008} convoluted with gsmooth. The best-fit spectrum is shown in Figure \ref{fig:zxipcf}. The broad positive Gaussian has a center energy of 6.72 keV, a width ($\sigma$) of 0.30 keV, and a normalization of $4.5\times10^{-3}$. The absorber is characterized by  $N_\mathrm{H}=2.8\times10^{21}$~cm$^{-2}$, $\log\xi=3.7$, and a velocity of $v = -0.02c$. It is broadened by gsmooth with a sigma of 40~eV, corresponding to $\sigma_v=2000$ km~s$^{-1}$.
This model reproduces the spectrum as well as the multiple-PIE model. It is difficult to statistically rule out either one with the current data. 
Future observations with richer photon statistic will be able to judge which model is preferred.

In this paper, we do not consider the absorption model further since the probability of an absorption line existing at precisely the right energy and depth to coincidentally split a single emission line is low.

It should be noted that the blueshifted absorber with $v=-0.02c$ would produce a \ion{Fe}{25} edge at 9.0 keV. However, as shown in Figure \ref{fig:zxipcf}, the depth of this edge is too small to affect the UFO feature.

\section{Robustness of the UFO line}
\begin{figure}[h]
   \centering
    \includegraphics[width=0.95\columnwidth]{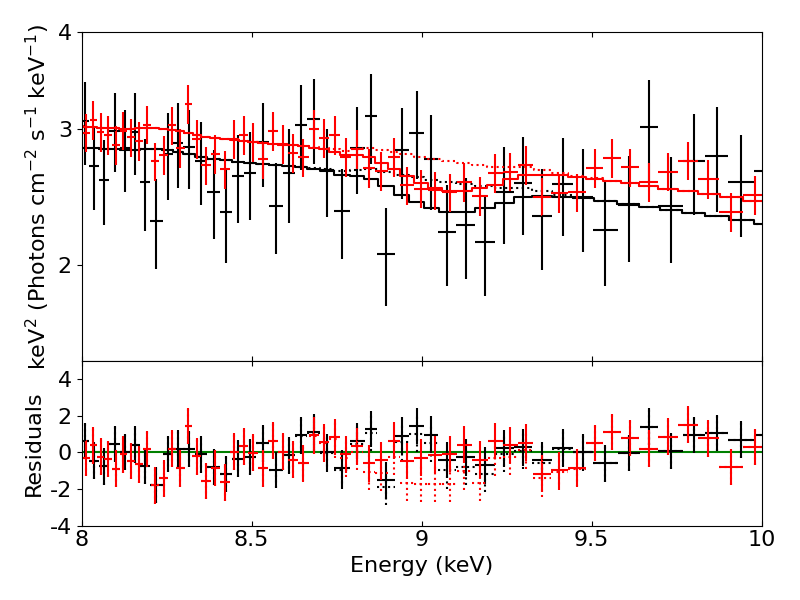}
            \caption{The Xtend (black) and Resolve (red) spectra in the 8--10~keV band. In the upper panel, the model line without/with the UFO absorption line is shown in the dotted/solid line, respectively. In the lower panel, the data bin show residuals in each model.}
         \label{fig:absline}
   \end{figure}

The UFO feature at 9 keV appears to be present in the Resolve spectrum ($G_3$), but it does not seem to be significant in Xtend (see also Figure \ref{fig:continuum}. Figure \ref{fig:absline} shows a zoomed-in view of the Xtend and Resolve spectra in the 8--10 keV band. In the case of Xtend, the spectrum can be explained equally well with or without a UFO component. Restricting the fit to 8--10 keV, the C-statistic changes only slightly, from 362.1 to 361.7 (by $-0.4$), when including the UFO. Therefore, while Xtend cannot detect the UFO feature, it also cannot rule out its presence.

\bibliography{ms}{}
\bibliographystyle{aasjournalv7}

\end{document}